\documentclass[fleqn]{article}
\usepackage{longtable}
\usepackage{graphicx}

\begin {document}
\begin{flushleft}
{\LARGE
{\bf Comment on ``Configuration interaction calculations and excitation rates of X-ray and EUV transitions in sulfurlike manganese" by El-Maaref et al.\,   [J. Elect.  Spectrosc. Related Phen.  215 (2017) 22]}
}\\

\vspace{1.5 cm}

{\bf {Kanti  M  ~Aggarwal}}\\ 

\vspace*{1.0cm}

Astrophysics Research Centre, School of Mathematics and Physics, Queen's University Belfast, \\Belfast BT7 1NN, Northern Ireland, UK\\ 
\vspace*{0.5 cm} 

e-mail: K.Aggarwal@qub.ac.uk \\

\vspace*{0.20cm}

Received: 25 April 2019; Accepted: 27 May 2019

\vspace*{1.0 cm}

\vspace*{1.0 cm}
\vspace*{1.0 cm}

\hrule

\vspace{0.5 cm}

\end{flushleft}

\clearpage


\begin{abstract}

In one of the papers, El-Maaref  et al.  [J. Elect.  Spectrosc. Related Phen.  215 (2017) 22] have reported results for energy levels, oscillator strengths (f-values), radiative rates (A-values), collision strengths ($\Omega$), and excitation rates for transitions in  S-like Mn~X. However, except for energy levels their results are restricted to only a few transitions and hence have limited application. Furthermore, most of their results have scope for improvement, but for $\Omega$ are not correct as these have been found to be overestimated by an order of magnitude. In this comment we discuss the discrepancies and deficiencies of their results and recommend that a fresh calculation should be performed for this ion. 

\end{abstract}

\clearpage

\section{Introduction}

In one of the papers, El-Maaref  et al. \cite{elm} have reported results for energy levels, oscillator strengths (f-values), radiative rates (A-values), collision strengths ($\Omega$), and excitation rates for transitions in  S-like Mn~X. They have considered 161 levels of the 3s$^2$3p$^4$, 3s3p$^5$, 3s$^2$3p$^3$3d, 3s$^2$3p$^3$4s/4p/4d, and 3s$^2$3p$^3$5s/5p configurations, i.e. 8 in total. For the calculations of energy levels they have adopted two different and independent atomic structure codes, namely the {\em configuration interaction version 3} (CIV3: \cite{civ3}) and the Los Alamos National Laboratory (LANL) code, which is mainly based on the Hartree-Fock relativistic code of Cowan \cite{cow}. Although they have listed energies for all the calculated levels, their corresponding results for f- and A-values are restricted to only a few transitions, while for $\Omega$ are presented  only graphically, and that too for 13 alone among the possible 12~880, i.e. $\sim$0.1\%. These limited data are not fully  sufficient for the modelling or analysis of plasmas. More importantly, differences between the two sets of energies are up to 3.3~eV for some levels (see their table~A), such as 38/39, whereas for the f-values the discrepancies are up to a factor of 20 for a few, see for example the transitions 3s$^2$3p$^4$~$^3$P$_1$ -- (3p$^3$($^2$D)4d) $^3$D$^o_1$ and $^3$P$^o_0$, which have significant magnitude, i.e. 0.60 and 0.37, respectively -- see their table~B. Such discrepancies clearly indicate that there is scope for improvement of accuracy in their presented results. However, our main concern is about their results for $\Omega$ which do not appear to be correct, particularly for the 1--57 (3p$^4$~$^3$P$_2$ -- 3p$^3$($^4$S)4p~$^5$P$_1$) transition -- see their fig.~1, because its value decreases sharply at the lower end of the energy range. Therefore, our focus in this comment is on the $\Omega$ results. 

\section {Energy levels and radiative rates}

In our calculations for all atomic parameters,   the Flexible Atomic Code (FAC: {\tt https://www-amdis.iaea.org/FAC/}) of Gu \cite{fac}  has been adopted. This is a fully relativistic code and is based on the many-body perturbation theory (MBPT) for the calculations of atomic structure and the {\em distorted wave} (DW) method for the scattering process. Furthermore, the code is highly efficient and reliable, and therefore is widely used by many workers in various parts of the world. We too have used it in the past for over a decade and for a wide range of ions from those of He to W. More importantly, it gives comparable results for most levels/transitions with other similar codes, such as CIV3 and GRASP, i.e. the General-purpose Relativistic Atomic Structure Package, available at the website {\tt http://amdpp.phys.strath.ac.uk/UK\_APAP/codes.html}, for energy levels and A-values, and the Dirac Atomic $R$-matrix Code (DARC), available at the same website as GRASP, for the collisional data. Therefore, we have confidence in our calculated results.

Since, as stated in section~1, our main interest is in the results for $\Omega$, in Table~1 we  list energies for only those levels which are required for the subsequent comparisons with the reported results of El-Maaref  et al. \cite{elm}. For consistency, we have included the same 161 levels (listed in section~1) and considered by El-Maaref  et al., and we refer to these results as FAC1. For the listed levels all energies agree to within $\sim$1~eV, which is highly satisfactory. Similar results for the f- and A-values for the relevant transitions are compared in Table~2. For some transitions the discrepancies are up to about a factor of two -- see for example 1--51/54, 2--51, and 3--51/54, for which the magnitudes are rather small, i.e. the transitions are comparatively weak. Such discrepancies among different calculations (for weak transitions) are often common and are understandable, because each code differs in methodology, algorithm, and/or the inclusion of relativistic effects. Furthermore, such transitions are more sensitive to {\em configuration interaction} (CI) than the stronger ones. However, two transitions in this table stand out, which are 1--90 and 3--92, both of which have f $\sim$ 0.13 in our calculations with FAC and those reported by El-Maaref  et al. with LANL, but their corresponding results with CIV3 are larger by about a factor of three, and do not appear to be correct. Nevertheless, since their subsequent calculations for $\Omega$ have been performed with the LANL code, also based on the DW method, it is satisfying to see a good agreement with our FAC  results, and it will be useful to discuss the comparisons for $\Omega$.

\begin{table}
\caption{Comparison of energies  (in eV) for some levels for S-like Mn~X. See El-Maaref et al. \cite{elm} for listing of all levels.} 
\begin{tabular}{rllrrrrrrrrrr} \hline
\\
Index &   Configuration & Level       &    CIV3    &   LANL  &   FAC1    &   FAC2   &   ADAS    \\ 
\hline \\												       
   1  & 3p$^4$            & $^3$P$_2$     &   0.00  &   0.00  &   0.00   &   0.00 &   0.00    \\
  2  & 3p$^4$            & $^3$P$_1$     &   1.22  &   1.20  &   1.23   &   1.22 &   1.06    \\
  3  & 3p$^4$            & $^3$P$_0$     &   1.49  &   1.44  &   1.50   &   1.50 &   1.34    \\
 48  & 3p$^3$($^4$S)4s   & $^3$S$^o_1$   & 119.56  & 119.75  & 119.36   & 121.35 & 129.43    \\
 51  & 3p$^3$($^2$D)4s   & $^3$D$^o_1$   & 123.71  & 123.19  & 122.89   & 124.64 & 132.65    \\
 54  & 3p$^3$($^2$P)4s   & $^3$P$^o_1$   & 126.68  & 127.10  & 127.09   & 128.93 & 136.51    \\
 57  & 3p$^3$($^4$S)4p   & $^5$P$_1$     & 129.95  & 129.58  & 128.69   & 130.45 & 138.48    \\
 90  & 3p$^3$($^4$S)4d   & $^3$D$^o_3$   & 149.50  & 148.94  & 148.28   & 150.21 & 159.40    \\
 91  & 3p$^3$($^4$S)4d   & $^3$D$^o_2$   & 149.56  & 148.88  & 148.21   & 150.10 & 159.31    \\
 92  & 3p$^3$($^4$S)4d   & $^3$D$^o_1$   & 149.60  & 148.93  & 148.26   & 150.11 & 159.51    \\
\\ \hline
\end{tabular} 

\begin{flushleft}
{\small
CIV3: earlier calculations of El-Maaref et al. \cite{elm} with  the {\sc civ3} code for 161 levels \\
LANL: earlier calculations of El-Maaref et al. \cite{elm} with  the {\sc lanl} code for 161 levels \\
FAC1:  present calculations  with  the {\sc fac} code for 161 levels \\
FAC2: present calculations with  the {\sc fac} code for 799 levels \\
ADAS:  earlier calculations of Alessandra Giunta with  the {\sc hfr} code for 799 levels, available at ADAS website {\tt http://open.adas.ac.uk/} \\
}
\end{flushleft}
\end{table}

\section {Collision strengths and effective collision strengths}

 El-Maaref  et al. \cite{elm} have shown their $\Omega$ results for 13 transitions, namely 1  (3p$^4$~$^3$P$_2$) to 48 (3p$^3$($^4$S)4s~$^3$S$^o_1$), 51 (3p$^3$($^2$D)4s~$^3$D$^o_1$), 54 (3p$^3$($^2$P)4s~$^3$P$^o_1$), 57 (3p$^3$($^4$S)4p~$^5$P$_1$), and 90 (3p$^3$($^4$S)4d~$^3$D$^o_3$); 2  (3p$^4$~$^3$P$_1$) to 48 (3p$^3$($^4$S)4s~$^3$S$^o_1$), 51 (3p$^3$($^2$D)4s~$^3$D$^o_1$), 54 (3p$^3$($^2$P)4s~$^3$P$^o_1$),  and 91 (3p$^3$($^4$S)4d~$^3$D$^o_2$); and 3  (3p$^4$~$^3$P$_0$) to 48 (3p$^3$($^4$S)4s~$^3$S$^o_1$), 51 (3p$^3$($^2$D)4s~$^3$D$^o_1$), 54 (3p$^3$($^2$P)4s~$^3$P$^o_1$),  and 92 (3p$^3$($^4$S)4d~$^3$D$^o_1$). We compare below their results with ours. 
 
 \subsection {Transitions from level 1} 
 
 Among the 5 transitions from the ground level shown by El-Maaref et al. \cite{elm} in their fig.~1, only 1--57 is forbidden and the rest are allowed. We show our results for the same 5 transitions in Fig.~1 (a and b). Unfortunately, no comparisons can be made with the earlier results, because El-Maaref et al. have not provided magnitude on the vertical scale, and therefore only the trends can be confirmed. As suspected, we do not agree with the behaviour of $\Omega$ shown by them for the 1--57 forbidden transition, for which their results at the lowest energy ($\sim$5~eV) are the {\em highest} among all the transitions shown in their fig.~1 at any energy, whereas ours are lower from the highest magnitude by more than a factor of 30, and neither is there  any  (known) reason for it to have values much higher than for the allowed ones. Therefore, their results for this transition are clearly wrong. 
 
Among the four allowed transitions the trends of increasing $\Omega$ with the increase of energy appear to be correct, except for 1--54 for which we observe a much higher magnitude towards the higher end of the energy range. For this transition the f-value ($\sim$ 0.01) is lower than for 1--48 or 1--90 (see Table~2) by about an order of magnitude, and so are the $\Omega$ results, as expected. So again, the results of El-Maaref et al. \cite{elm} do not appear to be correct for the 1--54 allowed transition. 

\begin{table}
\caption{Comparison of oscillator strengths (f-values) and radiative rates (A-values, s$^{-1}$)  for some transitions for S-like Mn~X. See Table~1 for definition of levels. a$\pm$b $\equiv$ a$\times$10$^{{\pm}\rm b}$.} 
\begin{tabular}{rrcccccccc} \hline
\\
I  & J &  f(CIV3)&   f (LANL) &     A(CIV3) &  f (FAC1)  & A(FAC1) & f (FAC2) & A (FAC2) &  A (ADAS)     \\ 
\hline \\												       
  1  & 48  & 7.66-2  & 1.06-1  & 7.91+10  & 7.64-2  & 7.87+10  & 8.11-2 & 8.63+10 & 7.68+10    \\ 
  1  & 51  & 9.91-4  & 1.51-3  & 1.10+09  & 1.36-3  & 1.49+09  & 7.30-4 & 8.20+08 & 1.09+09    \\
  1  & 54  & 1.33-2  & 1.01-2  & 1.55+10  & 8.08-3  & 9.43+09  & 1.03-2 & 1.24+10 & 1.06+10    \\
  1  & 90  & 3.67-1  & 1.33-1  & 2.54+11  & 1.21-1  & 8.21+10  & 1.31-1 & 9.16+10 & 6.02+10    \\
  2  & 48  & 7.79-2  & 8.34-2  & 4.74+10  & 6.21-2  & 3.76+10  & 6.44-2 & 4.03+10 & 3.63+10    \\
  2  & 51  & 2.52-2  & 4.70-2  & 1.64+10  & 3.47-2  & 2.23+10  & 1.89-2 & 1.25+10 & 1.71+10    \\
  2  & 54  & 1.31-2  & 1.74-2  & 8.98+09  & 1.32-2  & 9.08+09  & 1.63-2 & 1.16+10 & 9.78+09    \\
  2  & 91  & 3.05-1  & 8.92-2  & 1.75+11  & 8.56-2  & 4.82+10  & 8.39-2 & 4.84+10 & 1.68+09    \\
  3  & 48  & 7.54-2  & 9.41-2  & 1.52+10  & 6.98-2  & 1.40+10  & 7.26-2 & 1.51+10 & 1.32+10    \\
  3  & 51  & 9.77-2  & 7.57-2  & 2.11+10  & 5.69-2  & 1.21+10  & 3.12-2 & 6.84+09 & 1.02+10    \\
  3  & 54  & 4.94-2  & 1.25-1  & 1.12+10  & 8.96-2  & 2.04+10  & 1.09-1 & 2.56+10 & 2.17+10    \\
  3  & 92  & 3.67-1  & 1.36-1  & 1.16+10  & 1.29-1  & 4.02+10  & 4.90-2 & 1.56+10 & 1.72+10    \\
\\ \hline
\end{tabular} 

\begin{flushleft}
{\small
CIV3: earlier calculations of El-Maaref et al. \cite{elm} with  the {\sc civ3} code for 161 levels \\
LANL: earlier calculations of El-Maaref et al. \cite{elm} with  the {\sc lanl} code for 161 levels \\
FAC1:  present calculations  with  the {\sc fac} code for 161 levels \\
FAC2: present calculations with  the {\sc fac} code for 799 levels \\
ADAS:  earlier calculations of Alessandra Giunta with  the {\sc hfr} code for 799 levels, available at ADAS website {\tt http://open.adas.ac.uk/} \\
}
\end{flushleft}
\end{table}

 \begin{figure*}
\includegraphics[angle=-90,width=0.9\textwidth]{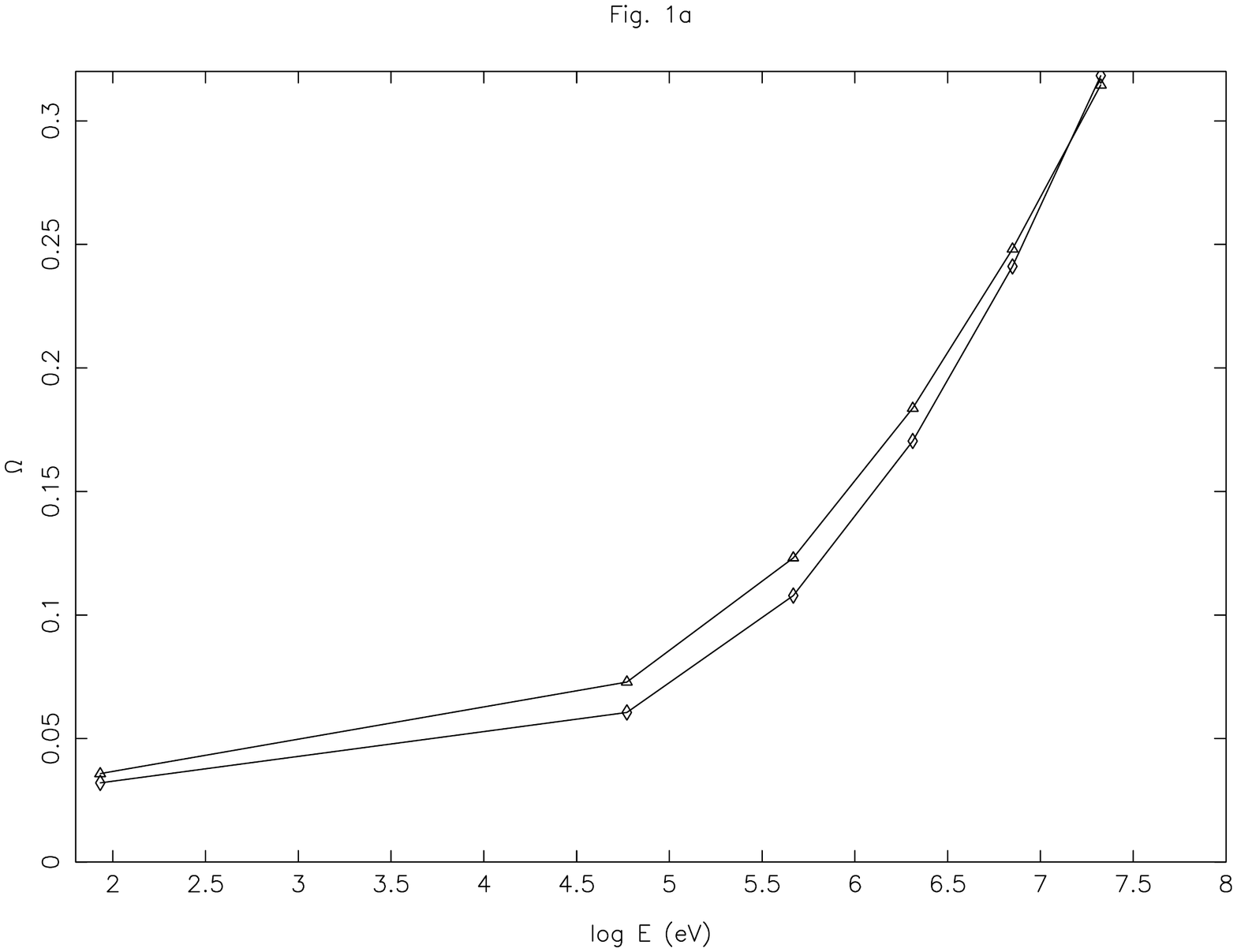}
 \vspace{-1.5cm}
 \end{figure*}
 
\setcounter{figure}{0}
 \begin{figure*}
\includegraphics[angle=-90,width=0.9\textwidth]{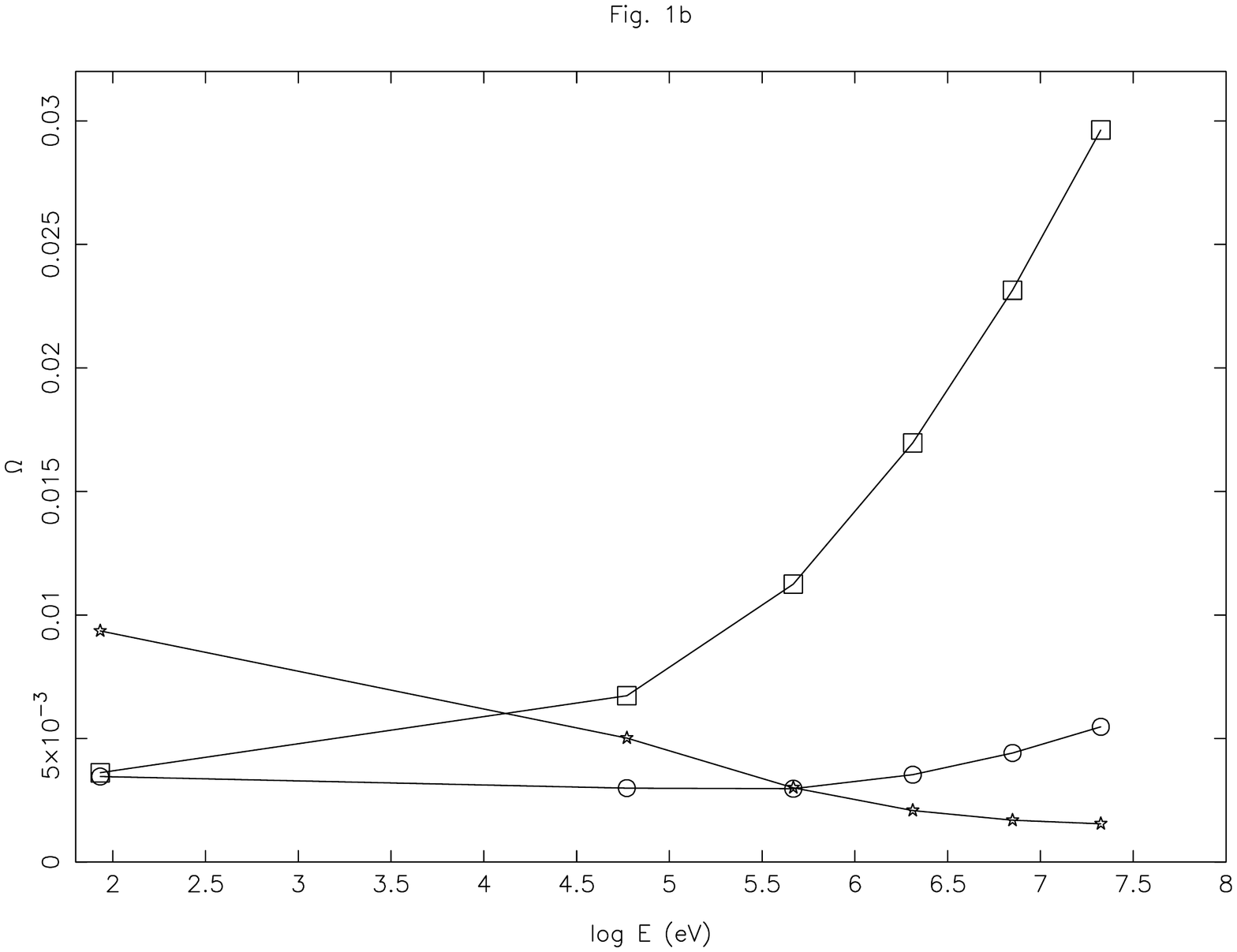}
 \vspace{-1.5cm}
 \caption{Collision strengths ($\Omega$) from  FAC  for a few transitions from the ground level of Mn~X,  (a) triangles: 1--48 (3p$^4$~$^3$P$_2$ -- 3p$^3$($^4$S)4s~$^3$S$^o_1$) and diamonds: 1--90 (3p$^4$~$^3$P$_2$ -- 3p$^3$($^4$S)4d~$^3$D$^o_3$), and (b) circles: 1--51 (3p$^4$~$^3$P$_2$ -- 3p$^3$($^2$D)4s~$^3$D$^o_1$), squares: 1--54 (3p$^4$~$^3$P$_2$ -- 3p$^3$($^2$P)4s~$^3$P$^o_1$), stars: 1--57 (3p$^4$~$^3$P$_2$ -- 3p$^3$($^4$S)4p~$^5$P$_1$).}
 \end{figure*}

 \subsection {Transitions from level 2}  
 
In Fig.~2 (a)  we show the variation of $\Omega$ with energy for 4 transitions from level 2 (2p$^4$~$^3$P$_1$) to higher excited ones, i.e. 48, 51, 54, and 91, the same as shown by El-Maaref et al. \cite{elm} in their fig.~2. Since they have provided magnitude on the vertical scale for these transitions, comparisons are possible with their results, as also shown in our Fig.~2. The only similarity in trends between the two sets of results is that $\Omega$ values are increasing with increasing energy, and this is wholly expected because all of them are allowed, but the magnitudes are  distinctly {\em different}, as their corresponding results are {\em higher} by up to an order of magnitude!
 
It may be noted that the x-scales in Figs. 1 and 2a are logarithmic, which is unconventional,  but has been adopted in order to facilitate direct comparisons with the corresponding results of El-Maaref et al. \cite{elm}. Such scales give the (false) impression that the $\Omega$ values rise fast with increasing energy, particularly towards the higher end. Therefore, in Fig.~2b we have shown the same results (and comparisons) again, but on a normal (linear) scale. Unfortunately, in this figure  our results from FAC appear to be abnormal. This is because our calculations cover a much wider range of energies, and therefore in yet another figure (Fig.~2c) we show $\Omega$ in the entire energy range. It becomes clear that the corresponding results of  El-Maaref et al. rise sharply towards the lower end, as these are falling equally sharply for the forbiddeen transition in their fig.~1.  This kind of behaviour (i.e. steep rise at very low energies) is normally observed for the `elastic' transitions, i.e. for which $\Delta$E is very small (nearly zero) -- see for example, fig.~4 of Aggarwal et al. \cite{fe26} for transitions of H-like Fe~XXVI. Therefore, we may conclude that neither the magnitudes nor the behaviours of their $\Omega$ data are correct for these transitions from level 2.

\setcounter{figure}{1}
\begin{figure*}
\includegraphics[angle=-90,width=0.9\textwidth]{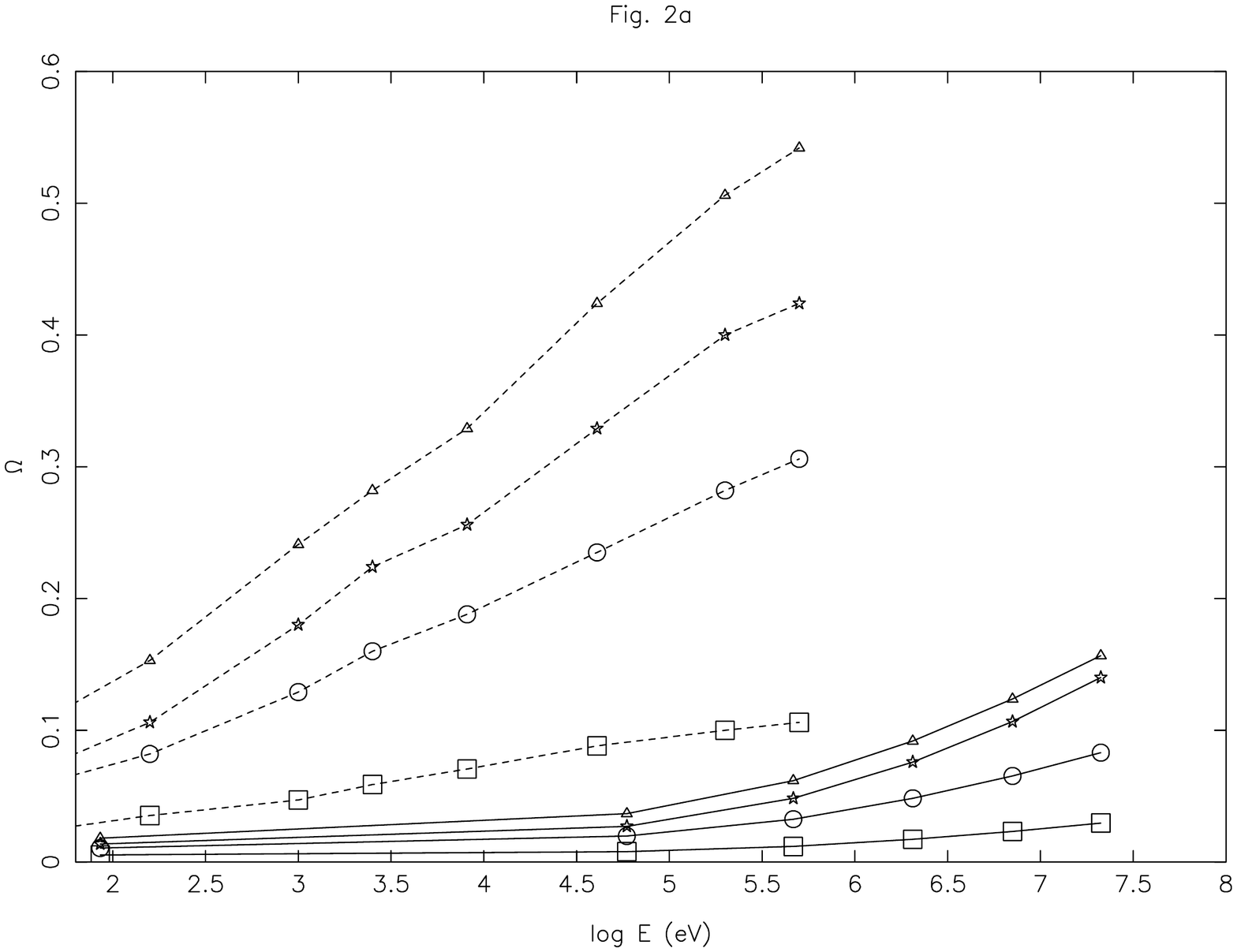}
 \vspace{-1.5cm}
  \end{figure*} 
  
 \setcounter{figure}{1}
\begin{figure*}
\includegraphics[angle=-90,width=0.9\textwidth]{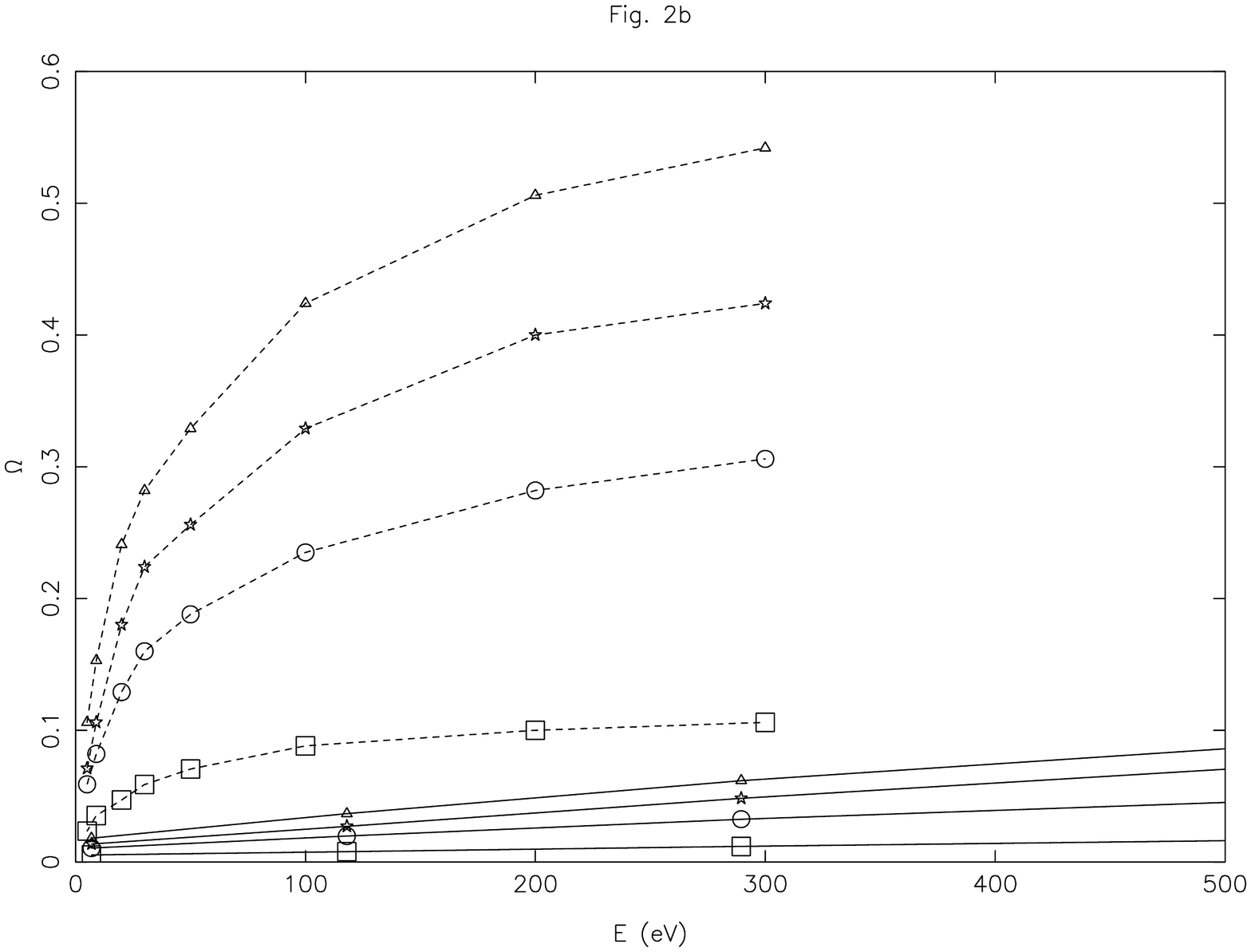}
 \vspace{-1.5cm}
 \end{figure*} 
 
\setcounter{figure}{1}
\begin{figure*}
\includegraphics[angle=-90,width=0.9\textwidth]{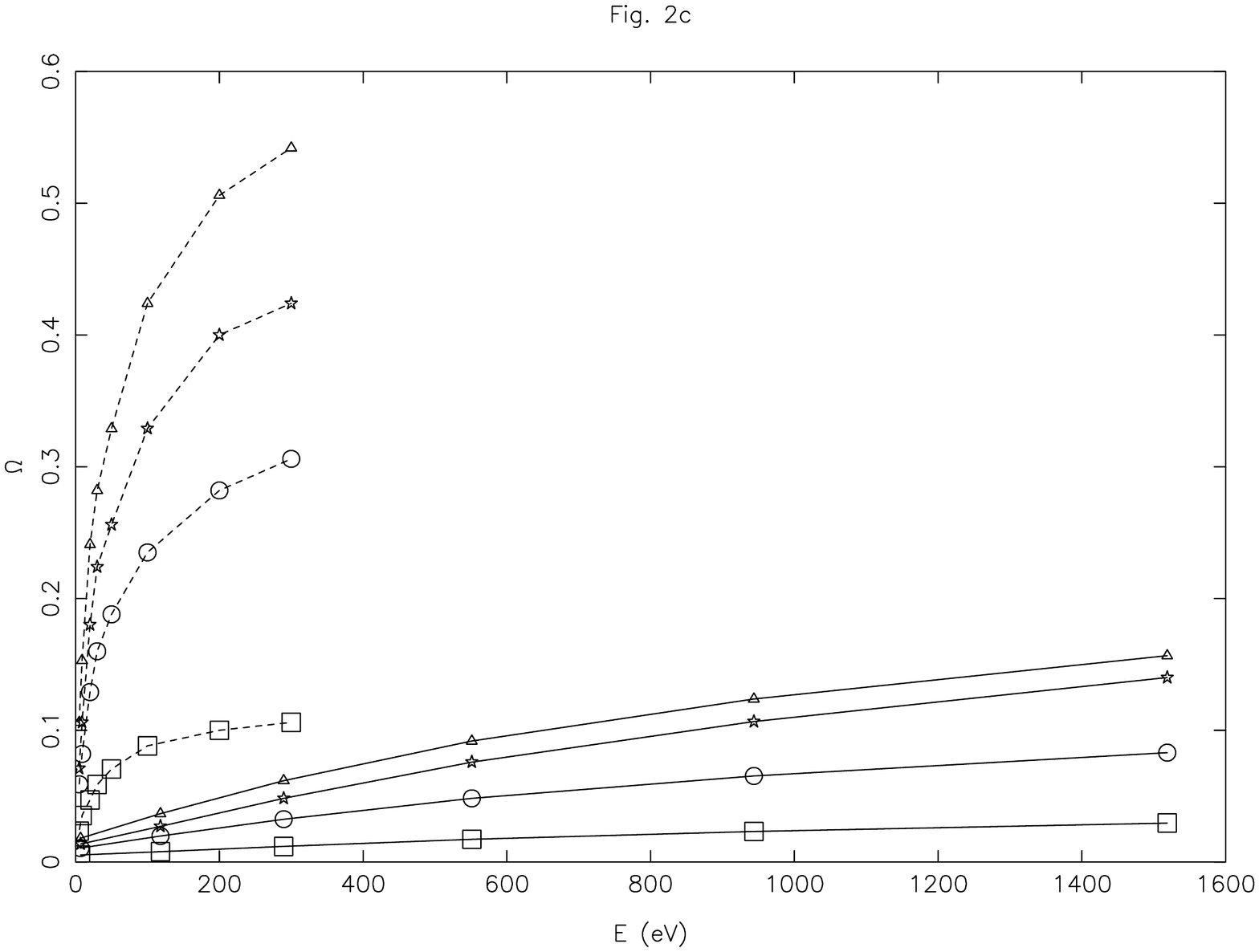}
 \vspace{-1.5cm}
\caption{Comparison of collision strengths ($\Omega$) from  FAC (continuous curves) and LANL (broken curves) codes  for a few transitions from the first excited level of Mn~X,  (a) triangles: 2--48 (3p$^4$~$^3$P$_1$ -- 3p$^3$($^4$S)4s~$^3$S$^o_1$),  circles: 2--51 (3p$^4$~$^3$P$_1$ -- 3p$^3$($^2$D)4s~$^3$D$^o_1$), squares: 2--54 (3p$^4$~$^3$P$_1$ -- 3p$^3$($^2$P)4s~$^3$P$^o_1$), and stars: 2--91 (3p$^4$~$^3$P$_1$ -- 3p$^3$($^4$S)4d~$^3$D$^o_2$). See text for  descriptions of (a), (b) and (c) panels.}
 \end{figure*} 
 
  \subsection {Transitions from level 3}  
  
In Fig.~3 we compare our results of $\Omega$ with those of  El-Maaref et al. \cite{elm} for 4 transitions from level 3 (2p$^4$~$^3$P$_0$) to 48, 51, 54 and 92, all of which are allowed. This comparison is similar, and so are the conclusions, as noted for transitions from level 2 in Fig.~2a. Therefore, it has become abundantly clear that the $\Omega$ results of El-Maaref et al. are different from ours  for all types of  transitions of Mn~X, and are larger by an order of magnitude. However, a reasonable question to ask by an unbiased reader is which one of the two calculations is more correct. Unfortunately there are no other results available in the literature with which to compare. This is because this ion has not attracted as much attention as its near neighbours, such as Fe ~XI and Ni~XIII, because not only its solar abundance is low but the spectrum is also not very rich. Fortunately however, unpublished data by Alessandra Giunta are available  at ADAS website {\tt http://open.adas.ac.uk/}. She has used the same {\sc hfr} code of Cowan \cite{cow} as El-Maaref et al. have for the calculations of energy levels and A-values. For collisional calculations DW method has been applied as in all other works. Furthermore, she has considered 799 levels belonging to the 3s$^2$3p$^4$, 3s3p$^5$, 3p$^6$, 3s$^2$3p$^3$3d, 3s3p$^4$3d, 3p$^5$3d, 3s3p$^3$3d$^2$, 3p$^4$3d$^2$, 3s$^2$3p$^3$4$\ell$, and 3s3p$^4$4$\ell$ configurations. However, results are available only for {\em effective} collision strengths ($\Upsilon$),  a parameter obtained after integration of $\Omega$ over an electron velocity distribution function, mostly {\em Maxwellian}, and required in the analysis or modelling of plasmas -- see eq. (8) of El-Maaref et al. Nevertheless, a comparison of our corresponding results for $\Upsilon$ with theirs will be highly useful for the further confirmation of the (in)accuracy of our calculations for $\Omega$. 
  
\setcounter{figure}{2} 
\begin{figure*}
\includegraphics[angle=-90,width=0.9\textwidth]{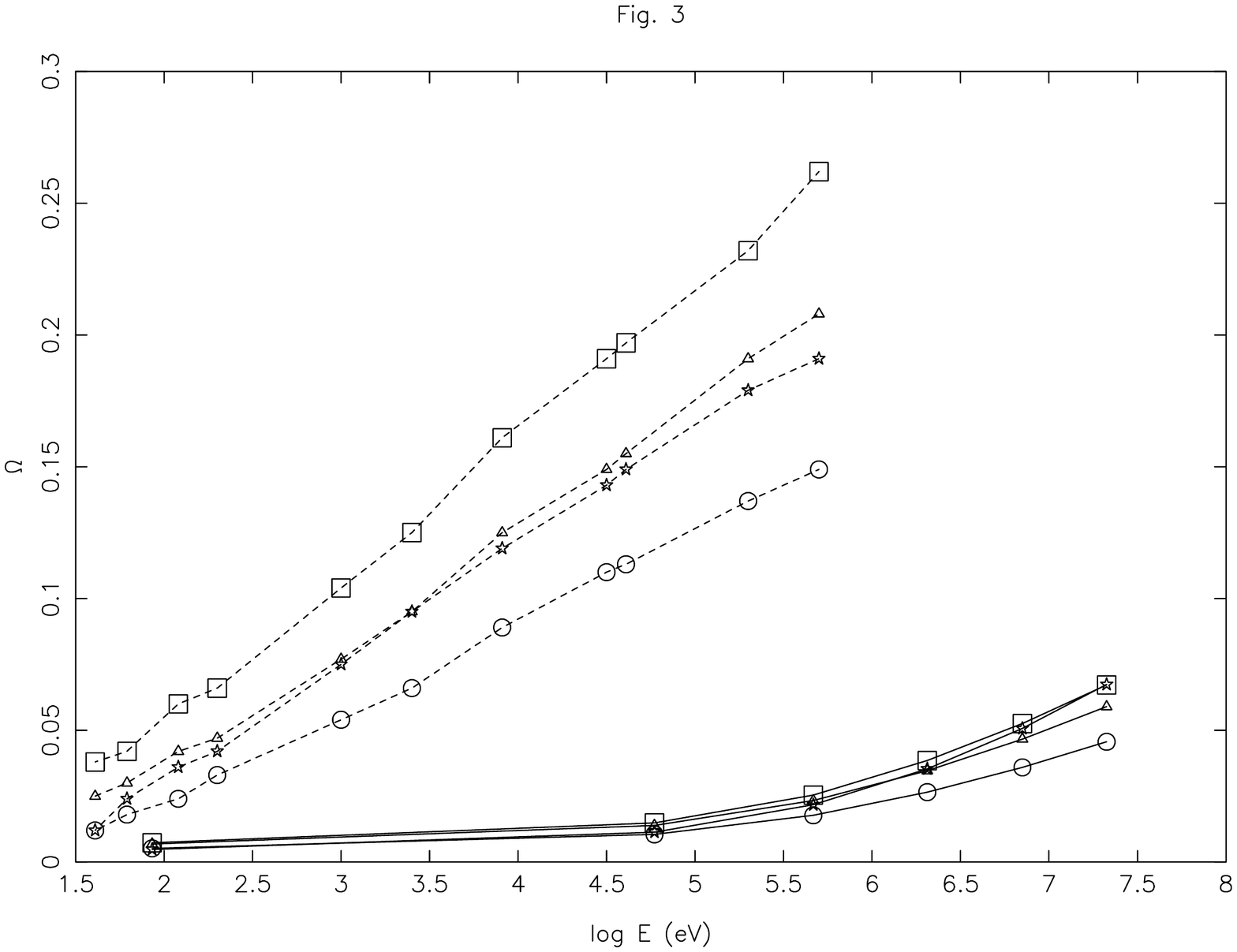}
 \vspace{-1.5cm}
\caption{Comparison of collision strengths ($\Omega$) from  FAC (continuous curves) and LANL (broken curves) codes  for a few transitions from the second excited level of Mn~X,  (a) triangles: 3--48 (3p$^4$~$^3$P$_0$ -- 3p$^3$($^4$S)4s~$^3$S$^o_1$),  circles: 3--51 (3p$^4$~$^3$P$_0$ -- 3p$^3$($^2$D)4s~$^3$D$^o_1$), squares: 3--54 (3p$^4$~$^3$P$_0$ -- 3p$^3$($^2$P)4s~$^3$P$^o_1$), and stars: 3--92 (3p$^4$~$^3$P$_0$ -- 3p$^3$($^4$S)4d~$^3$D$^o_1$).}
 \end{figure*}
 
 \subsection {Effective collision strengths} 
 
 Before we discuss our results for $\Upsilon$, we will like to note that we have also performed a larger calculation with FAC with the same 799 levels (FAC2), and the determined energies, f- and A-values are included in Tables~1 and 2 for a ready comparison, and so are the results from ADAS. Considering that CI between FAC1 and FAC2 calculations is very different, the agreement between the two is quite reasonable as there are no large discrepancies. Similarly, there are no large discrepancies with the corresponding results of ADAS, although looking at Table~1 one may infer that these results are comparatively  less accurate, because energy differences with the rest of the work are up to about 10~eV. However, our aim here is not to discuss such minor discrepancies but the larger ones, seen in Figs.~2 and 3. Since $\Omega$ values for allowed transitions mainly depend on the strengths of f and their $\Delta$E, there should be a satisfactory agreement between our calculations for $\Upsilon$ and those at ADAS, particularly for levels with the 3p$^3$4s configuration. 

In Fig.~4 (a and b) we compare our values of $\Upsilon$ with those of ADAS for 4 transitions, namely 1--48 (3p$^4$~$^3$P$_2$ -- 3p$^3$($^4$S)4s~$^3$S$^o_1$), 1--51 (3p$^4$~$^3$P$_2$ -- 3p$^3$($^2$D)4s~$^3$D$^o_1$),  1--54 (3p$^4$~$^3$P$_2$ -- 3p$^3$($^2$P)4s~$^3$P$^o_1$), and 1--57 (3p$^4$~$^3$P$_2$ -- 3p$^3$($^4$S)4p~$^5$P$_1$),  out of which only the last one is forbidden. However, there are no significant discrepancies between the two sets of results, and throughout the whole temperature range. Similar comparisons are shown in Fig.~5 for the 2--48 (3p$^4$~$^3$P$_1$ -- 3p$^3$($^4$S)4s~$^3$S$^o_1$),   2--51 (3p$^4$~$^3$P$_1$ -- 3p$^3$($^2$D)4s~$^3$D$^o_1$), and  2--54 (3p$^4$~$^3$P$_1$ -- 3p$^3$($^2$P)4s~$^3$P$^o_1$) transitions, all of which are allowed. Finally, comparisons are shown in Fig.~6 for three allowed transitions from level 3, i.e. 3--48 (3p$^4$~$^3$P$_0$ -- 3p$^3$($^4$S)4s~$^3$S$^o_1$),   3--51 (3p$^4$~$^3$P$_0$ -- 3p$^3$($^2$D)4s~$^3$D$^o_1$), and 3--54 (3p$^4$~$^3$P$_0$ -- 3p$^3$($^2$P)4s~$^3$P$^o_1$), and again, there are no appreciable discrepancies. A good agreement of $\Upsilon$ values shown in Figs.~4--6 for 11 transitions, in spite of having different levels of CI, {\em confirms} that our results for $\Omega$ (and subsequently of $\Upsilon$) are {\em correct}, whereas those of El-Maaref et al. \cite{elm} are not.

\setcounter{figure}{3} 
 \begin{figure*}
\includegraphics[angle=-90,width=0.9\textwidth]{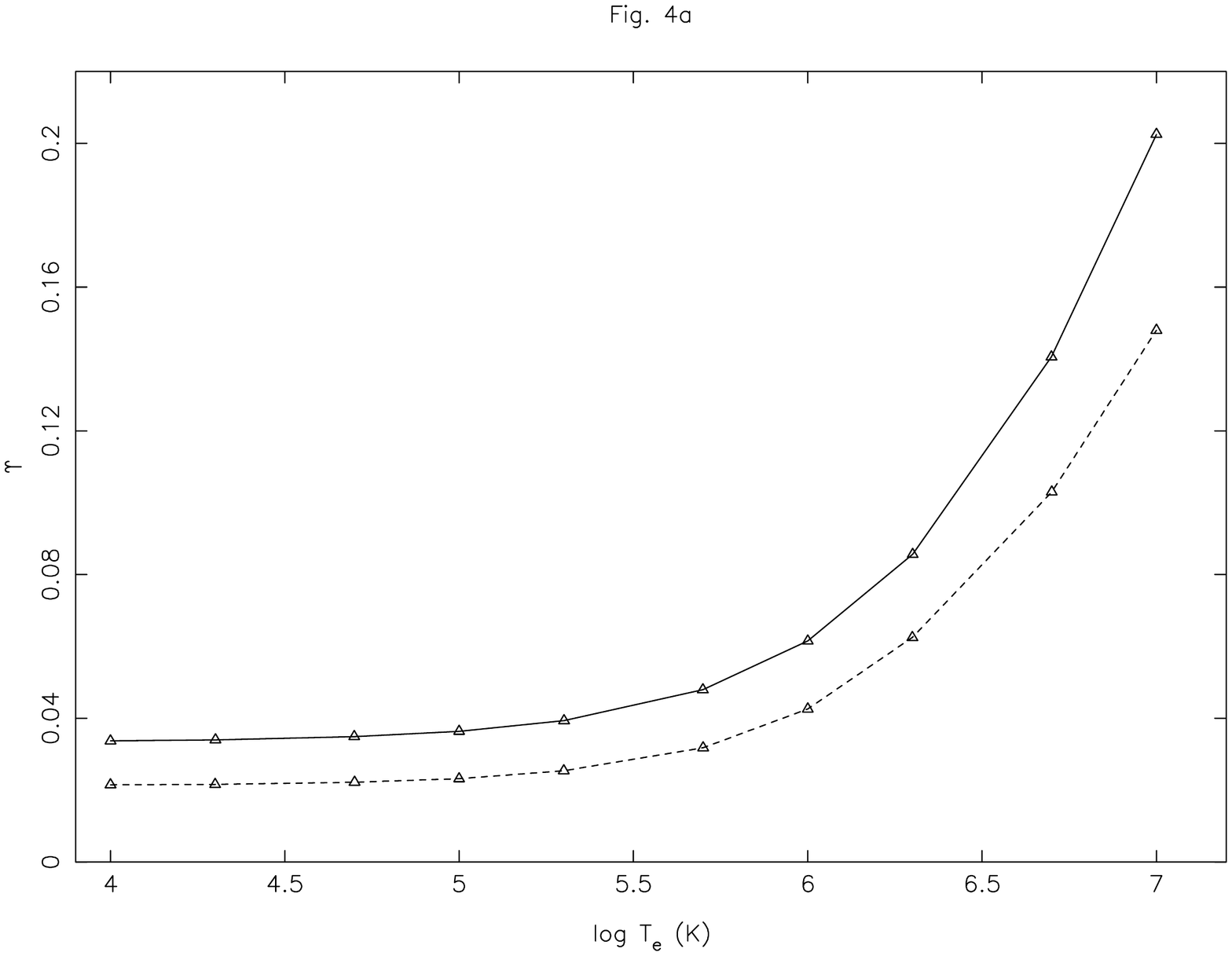}
 \vspace{-1.5cm}
 \end{figure*}
 
\setcounter{figure}{3}
 \begin{figure*}
\includegraphics[angle=-90,width=0.9\textwidth]{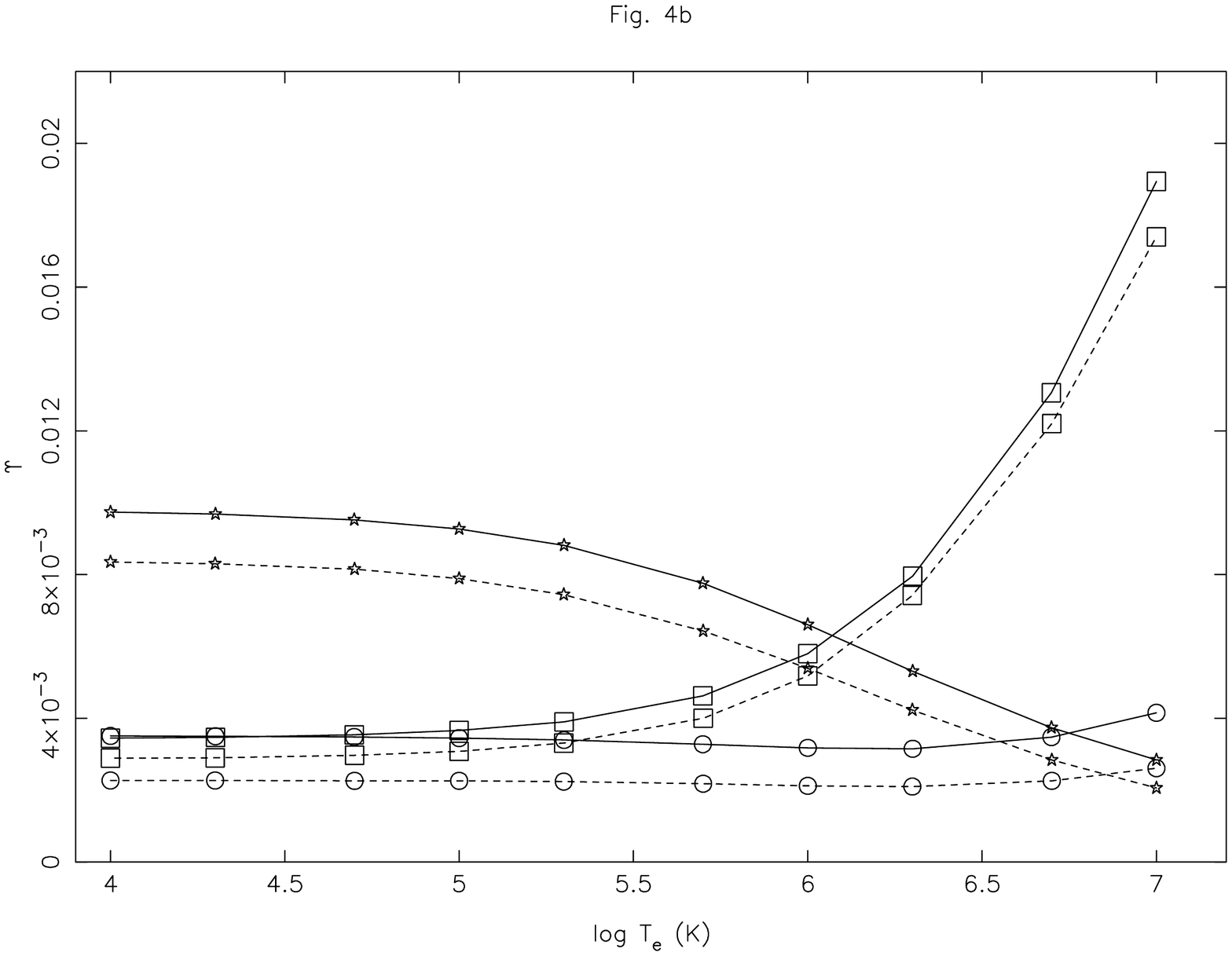}
 \vspace{-1.5cm}
 \caption{Comparison of effective collision strengths ($\Upsilon$) from  FAC  (continuous curves) and DW (broken curves) codes for a few transitions from the ground level of Mn~X,  (a) triangles: 1--48 (3p$^4$~$^3$P$_2$ -- 3p$^3$($^4$S)4s~$^3$S$^o_1$), and (b) circles: 1--51 (3p$^4$~$^3$P$_2$ -- 3p$^3$($^2$D)4s~$^3$D$^o_1$), squares: 1--54 (3p$^4$~$^3$P$_2$ -- 3p$^3$($^2$P)4s~$^3$P$^o_1$), and stars: 1--57 (3p$^4$~$^3$P$_2$ -- 3p$^3$($^4$S)4p~$^5$P$_1$).}
 \end{figure*}

\setcounter{figure}{4}
 \begin{figure*}
\includegraphics[angle=-90,width=0.9\textwidth]{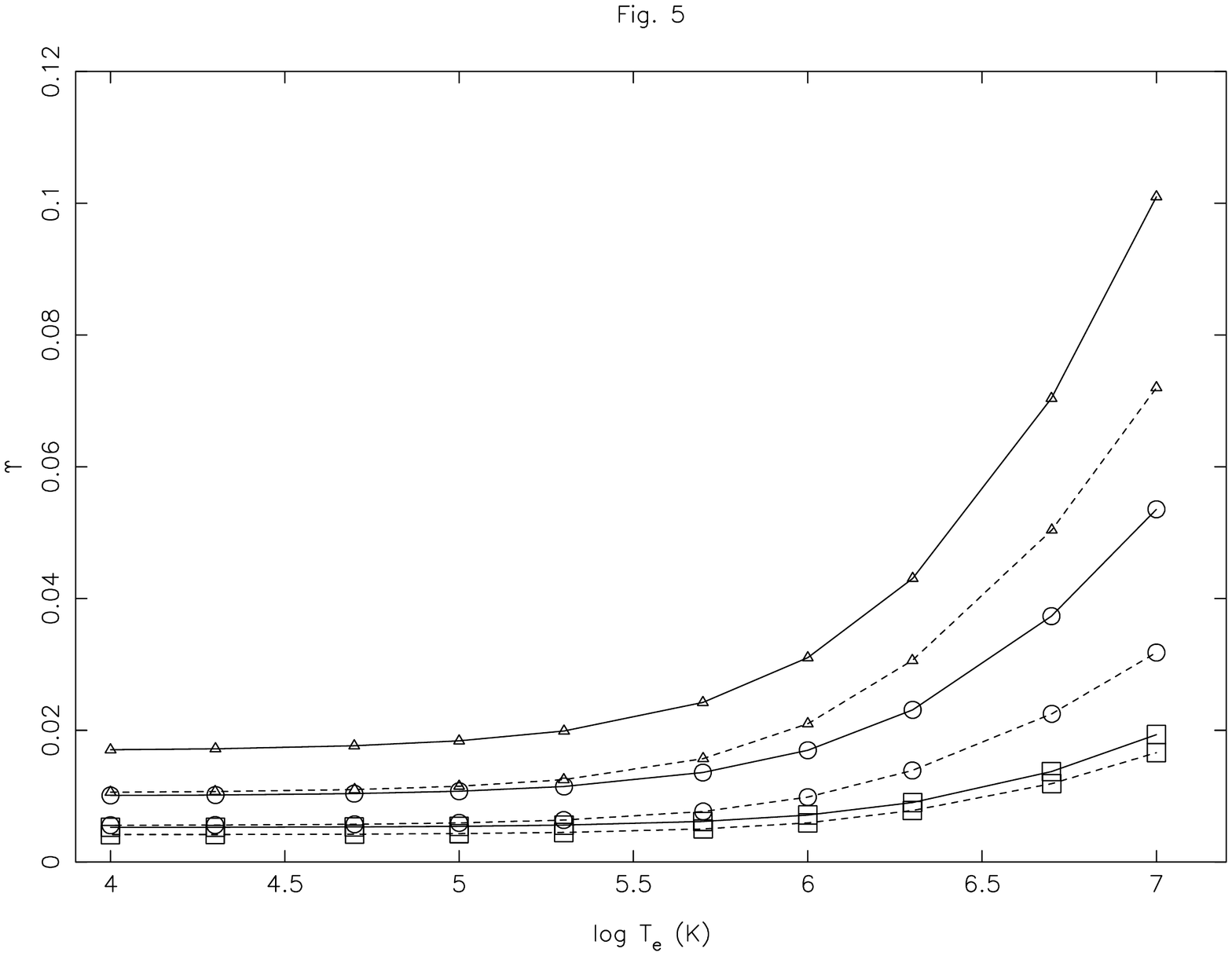}
 \vspace{-1.5cm}
 \caption{Comparison of effective collision strengths ($\Upsilon$) from  FAC  (continuous curves) and DW (broken curves) codes  for a few transitions from the first excited level of Mn~X,   triangles: 2--48 (3p$^4$~$^3$P$_1$ -- 3p$^3$($^4$S)4s~$^3$S$^o_1$),  circles: 2--51 (3p$^4$~$^3$P$_1$ -- 3p$^3$($^2$D)4s~$^3$D$^o_1$), and squares: 2--54 (3p$^4$~$^3$P$_1$ -- 3p$^3$($^2$P)4s~$^3$P$^o_1$).}
 \end{figure*} 
 
 \setcounter{figure}{5}
 \begin{figure*}
\includegraphics[angle=-90,width=0.9\textwidth]{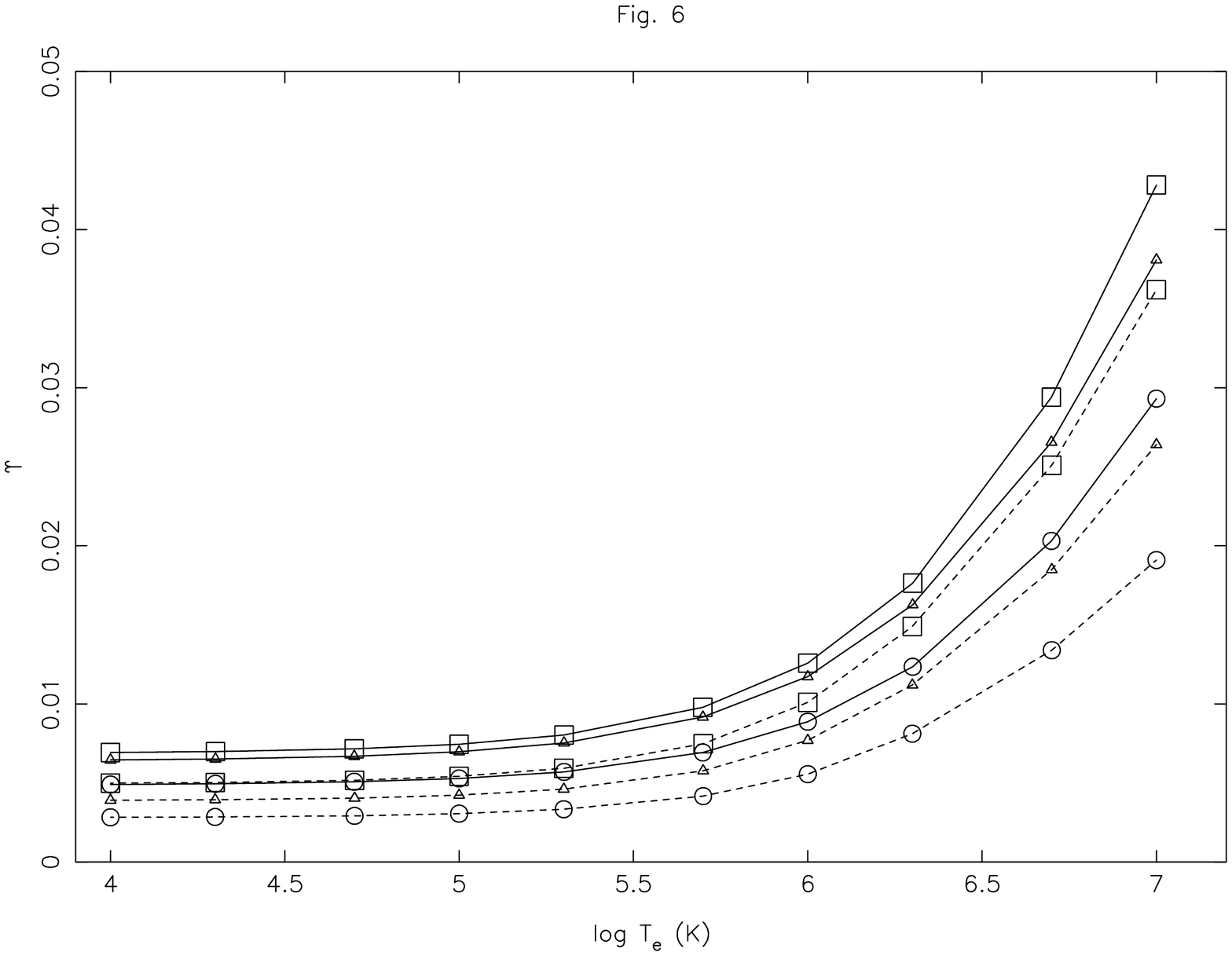}
 \vspace{-1.5cm}
 \caption{Comparison of effective collision strengths ($\Upsilon$) from  FAC  (continuous curves) and DW (broken curves) codes for a few transitions from the second excited level of Mn~X,  triangles: 3--48 (3p$^4$~$^3$P$_0$ -- 3p$^3$($^4$S)4s~$^3$S$^o_1$),  circles: 3--51 (3p$^4$~$^3$P$_0$ -- 3p$^3$($^2$D)4s~$^3$D$^o_1$), and squares: 3--54 (3p$^4$~$^3$P$_0$ -- 3p$^3$($^2$P)4s~$^3$P$^o_1$).}
 \end{figure*}

\section{Conclusions}

In this comment, through our independent calculations with FAC, we have demonstrated that while the earlier reported results of  El-Maaref et al. \cite{elm}  for Mn~X have scope for improvement for energy levels and A-values, their corresponding calculations for $\Omega$ are incorrect, in both behaviour and magnitude. This has been further confirmed by the comparisons made for $\Upsilon$ with the earlier available (although unpublished) results on the ADAS website. 

Mn~X is not a very important ion but still has applications in both fusion and astrophysical plasmas. The results of El-Maaref et al. \cite{elm} for several atomic parameters are very limited apart from being deficient and unreliable. For the determination of energy levels and A-values, much larger calculations involving a significantly more CI are required to achieve a good accuracy. Similarly, for the calculations of $\Omega$ more sophisticated methods/codes (such as DARC) need to be engaged, because these will elucidate the closed channel (Feshbach) resonances in the thresholds regions, whose contributions in the determinations of $\Upsilon$ are often dominant and highly significant, particularly for the forbidden transitions.  However, such calculations are often very demanding in terms of manpower and computational resources. In the meantime, if required, simple calculations particularly with FAC can be easily performed. Alternatively, results can be obtained from the author on request or can be downloaded from the ADAS website. Finally, for the benefit of readers we will like to note that the earlier results of El-Maaref et al. reported for other ions (W~XXXIX and W~XLV) are also incorrect, as recently demonstrated and explained by us \cite{w39},\cite{w45}.

\section*{Acknowledgment}  

We thank Dr. Martin O'Mullane for helpful correspondence regarding the ADAS data.

\end{document}